# ChatGPT and Vaccine Hesitancy: A Comparison of English, Spanish, and French Responses Using a Validated Scale

Saubhagya Joshi, Eunbin Ha, MA, Yonaira Rivera, PhD, and Vivek K. Singh, PhD
School of Communication & Information, Rutgers University, New Brunswick, NJ, USA.

**Abstract**

*ChatGPT is a popular information system (over 1 billion visits in August 2023) that can generate natural language responses to user queries. It is important to study the quality and equity of its responses on health-related topics, such as vaccination, as they may influence public health decision-making. We use the Vaccine Hesitancy Scale (VHS) proposed by Shapiro et al.[1] to measure the hesitancy of ChatGPT responses in English, Spanish, and French. We find that: (a) ChatGPT responses indicate less hesitancy than those reported for human respondents in past literature; (b) ChatGPT responses vary significantly across languages, with English responses being the most hesitant on average and Spanish being the least; (c) ChatGPT responses are largely consistent across different model parameters but show some variations across the scale factors (vaccine competency, risk). Results have implications for researchers interested in evaluating and improving the quality and equity of health-related web information.*

**Introduction**

Information systems like Google Search and ChatGPT are becoming informational infrastructures influencing public health[2,3] and there are calls to consider information as an important determinant of health[4]. Globally, over 100 million daily searches relevant to health are carried out on Google and global views of YouTube videos related to health surpassed a staggering 110 billion times in 2021[5,6]. Similarly, ChatGPT has recently been recording over a billion visits on a monthly basis[7] and is increasingly being used for medical applications[8]. ChatGPT can assist by having a direct conversation with patients and offering them quick and easy health support tailored to their needs. Previous studies have shown that ChatGPT can provide expert-level insights and logical reasoning in specific health contexts, highlighting its potential for public health applications[9,10]. At the same time, other researchers have reported concerns about misinformation and bias in ChatGPT responses[11] and the limitations when using ChatGPT for healthcare purposes[2].

Vaccine hesitancy refers to a delay in acceptance or refusal of vaccines despite the availability of vaccination services and supporting evidence[12]. Several studies have shown that online information can affect people's attitudes, beliefs, and behaviors regarding vaccination, both positively and negatively[13]. Unfortunately, vaccine effectiveness and the associated risks remain hotly contested at a societal level despite overwhelming scientific evidence and consensus supporting their safety and significance[14]. Multiple studies have reported on misleading vaccination-related information resources, such as websites, social media posts, leaflets, and publications, and their negative impacts on vaccine adoption[15,16]. Therefore, it is essential to evaluate the quality, especially in terms of congruence to scientific consensus, as emanated by newer online information resources like ChatGPT that are rapidly being adopted by end users for health-related queries[8].

At the same time, there exist important disparities in vaccination attitudes and vaccination rates across different sections of society. A recent systematic review[17] found inequality in the distribution of COVID-19 vaccines between and within countries. Hence, it would be important to ensure that emerging information systems (e.g., ChatGPT) do not exacerbate such inequities and provide parity in health-related information for different sections of society (e.g., those with different primary language preferences). Notably, earlier work has revealed that natural language processing (NLP) datasets and large language models (LLMs) perform better for English-speaking populations and show a potential disparity between English speakers and non-native English groups[18,19]. ChatGPT builds upon large language models and recent experimental research on ChatGPT's performance including 37 diverse languages has shown persistent English-centric biases of ChatGPT[20]. The dominance of English content in training data can be problematic since non-English populations can be underrepresented in datasets. A recent explanation for this process is that even though ChatGPT builds upon models that are multilingual, the languages don't necessarily inform one another[21]. Further, variations in the language of interaction can have different persuasive effects on the population. A recent study has revealed that when COVID-19 vaccine information was presented in English rather than the native language (Chinese), Hongkongers showed a higher level of confidence in the vaccine's safety and efficacy[22]. Despite the potential impact of language on individual health decisions, little is known about language-based differences in ChatGPT responses regarding vaccine hesitancy. However, these differences across languages can compromise health equity and are an important challenge for the field of public health[23].

Taken together, the above trends motivate more research that focuses on the **quality** and **equity** of health content generated by AI-powered tools such as ChatGPT, especially in contested domains such as vaccination.

We are particularly interested in exploring the social and technological ramifications of ChatGPT's responses to vaccine hesitancy-related queries. From a social perspective, it is important to identify how ChatGPT responds to ideologically divisive topics such as vaccination. While concerns about vaccine hesitancy have regained public attention due to the implementation of COVID-19 vaccinations, vaccine hesitancy is an issue that predates the pandemic[24,25]. Reasons for hesitancy span from lack of trust in vaccines and institutions to belief-based extremism, with additional concerns related to how digital media amplifies vaccine questioning[25]. Further, considering digital equity as an important determinant of health, uncovering any differences that exist in how ChatGPT responds to questions related to vaccine hesitancy can shed light on important disparities in the responses that the audiences may be exposed to when conversing with ChatGPT about health-related recommendations. Responses based on information that is not consistent with scientific consensus may further impact hesitancy and inequities among vulnerable populations, such as individuals who do not have a primary care provider to answer their questions about vaccine safety and effectiveness. From a technological perspective, it is also important to assess the equity of responses across languages in such an emerging health-relevant information system as ChatGPT. The same health-related questions posed in different languages should yield the same answers. Otherwise, there is a risk of unequal access of different demographic groups to health information for their interactions with such systems.

Therefore, in this study, language-specific differences in vaccine hesitancy-related responses generated by ChatGPT are discussed in light of recent concerns about information quality and potential disparities based on language used in information systems. The specific research questions of this study are as follows:
*(RQ1) How do ChatGPT responses regarding vaccine hesitancy compare with human population responses as reported in prior research?*
*(RQ2) What are the similarities and differences in ChatGPT responses to vaccine hesitancy when queried in English, Spanish, and French?*
*(RQ3) How are the variations in responses (if any) associated with different factors (competence of vaccines, risk) of the vaccine hesitancy scale and different technical parameters of the underlying models?*

**Methods**

*Study Design*
To conduct the study, we employed the Vaccine Hesitancy Scale (VHS), a validated scale used in previous survey research by Shapiro et al.[1]. The VHS used nine items to assess the attitudes of parents regarding lack of confidence in vaccines and the risk associated with vaccinating their children through a 5-point Likert scale. A survey was selected as a measure of assessment because we wanted to effectively compare variations in the responses of ChatGPT to a standard scale but using different languages. It also serves as a proxy for inferring how ChatGPT is likely to respond to other vaccine-related questions posed by different users and provides clues on how the LLMs underlying ChatGPT may be learning from the training data in various languages regarding vaccine hesitancy.

Items from Shapiro's VHS were used to collect responses in three languages: English (EN), Spanish (ES), and French (FR). First, we collected VHS responses for English and French languages and compared our results to results coming from human respondents as reported in Shapiro et al.[1]. Then, we extended our investigation to Spanish since it is the most widely used language in the US after English according to the US Census Bureau[26].

Following Shapiro et al.[1], we also analyzed variation in the VHS factors measuring lack of confidence in vaccines and perceived vaccine risks ('lack of confidence' and 'risk'). Questions 5 and 9 were used to quantify 'risk' and the remaining questions (1-4, 6-8) were used for 'lack of confidence'. The responses to 'lack of confidence' questions were reverse coded, such that a 'Strongly Agree' denotes the lowest value (1), i.e., the least vaccine hesitancy.

*Data Collection*

We used the OpenAI API[27] to interface with the latest publicly available iterations (GPT-4-0613 and GPT-3.5-Turbo-0613)[28] of the LLM implementations of OpenAI. During the first week of September 2023, we collected data through the chat completion endpoint (https://api.openai.com/v1/chat/completions)[29]. Our approach was to collect data systematically to support comparisons. One configuration was used as the "primary" model and we systematically varied three parameters (model of GPT engine, all questions in one session or one-by-one, default temperature) one at a time to create "variation" models for comparison across models and for overall trend analysis.

The choice of primary model (GPT-3.5-Turbo, which we call *turbo* here onward) was based on the currently available ChatGPT engine for OpenAI free-tier users. Paying users also have access to the GPT-4 engine. GPT-4 can follow more complex and longer instructions in the chat context and is typically more accurate than *turbo*. But *turbo* is faster and better suited to simple instructions in chat format. For our implementation, we used GPT-4 as a variation from our primary model.

Temperature is an optional parameter for chat completions that allows users to instruct ChatGPT how deterministic they would like their responses to be. (Lower values of temperature lead to more deterministic answers.) By default, the value of temperature is set to 1 for chat completions[30]. Possible values for temperature range from 0 to 2 in the current API. The variation model in our implementation used a temperature setting of 0.5.

We define a session as one GPT chat instance including the context prompts, chat overhead, and the response. The primary model included 9 VHS questions per session. This primary model mimics the setting used by Shapiro et al.[1], where the relevant randomized survey questions were filled by a respondent in a session. The variation model included one question per session. The settings for different models considered are summarized in Table 1.

**Table 1.** Model configurations used in the study

|  | **Primary Model** | **Variations** (Differences Only) |
|---|---|---|
| **GPT-Engine** | GPT-3.5-Turbo-0613 | GPT-4-0613 |
| **Temperature** | (Default) | 0.5 |
| **Questions** | 9 VHS questions per session | 1 VHS question per session |
| **Languages** | EN, ES, & FR | - |
| **Runs** | 30 sessions x 3 languages | 9 questions x 30 sessions x 3 languages |

Table 2 shows the prompts used to interact with ChatGPT. The English and French VHS questions are from the final validated scale in Shapiro et al.[1]. The Spanish translations were validated by a bilingual, bicultural team member who is an expert in public health communication and health inequities among Latinx and underserved communities.

The Python code used to collect data consisted of a "manager" that runs the code according to a configuration file, the GPT interface, and the "encoder" that codes responses into measurable values according to the VHS. The GitHub repository is available at: GitHub - Behavioral-Informatics-Lab/ChatGPT-language-bias.

**Data Analysis**

Data was analyzed by using the Analysis ToolPak add-in of Microsoft Excel for Microsoft 365 MSO. First, a one-way ANOVA was conducted to test if there were any differences in the VHS means by language. Thereafter, post hoc t-tests assuming unequal variances were conducted to examine which language differences were significant. Second, a two-way ANOVA was conducted to examine differences in VHS means across languages (EN, ES, and FR) and the VHS factors ('lack of confidence' and 'risk'). Follow-up post hoc pairwise t-tests between languages, assuming unequal variances, were conducted for each factor. These tests were conducted for the primary model as well as for configurations with variations.

**Table 2**. Prompt structure for each language*

| Language | Prompt |
|---|---|
| English | How much do you agree with each of the following statements on vaccinations? Please give me only one answer for each statement: Strongly disagree, Disagree, Neither agree or disagree, Agree, Strongly agree<br><br>(1) Childhood vaccines are important for my child's health; (2) Childhood vaccines are effective; (3) Having my child vaccinated is important for the health of others in my community; (4) All childhood vaccines offered by the government program in my community are beneficial; (5) New vaccines carry more risks than older vaccines; (6) The information I receive about vaccines from the vaccine program is reliable and trustworthy; (7) Getting vaccines is a good way to protect my child from disease; (8) Generally I do what my doctor or health care provider recommends about vaccines for my child; (9) I am concerned about serious adverse effects of vaccines |
| Spanish | ¿Qué tan de acuerdo está con cada una de las siguientes afirmaciones sobre las vacunas? Por favor, dame una sola respuesta para cada afirmación: Totalmente de acuerdo, De acuerdo, Ni de acuerdo ni en desacuerdo, En desacuerdo, Totalmente en desacuerdo<br><br>(1) Las vacunas para niños son importantes para la salud de mi hijo; (2) Las vacunas para niños son efectivas; (3) Tener a mi hijo vacunado es importante para la salud de los demás en mi comunidad; (4) Todas las vacunas para niños que ofrece el programa gubernamental en mi comunidad son beneficiosas; (5) Las vacunas nuevas conllevan más riesgos que las vacunas más antiguas; (6) La información que recibo sobre las vacunas del programa de vacunas es confiable y fidedigna; (7) Vacunarse es una buena manera de proteger a mi hijo de enfermedades; (8) En general, hago lo que recomienda mi médico o proveedor de atención médica sobre las vacunas para mi hijo; (9) Me preocupan los efectos adversos graves de las vacunas |
| French | Dans quelle mesure êtes-vous d'accord avec chacune des affirmations suivantes concernant les vaccins? Veuillez me donner une seule réponse pour chaque affirmation: Tout à fait d'accord, D'accord, Ni d'accord ni en désaccord, En désaccord, Fortement en désaccord<br><br>(1) Les vaccins pour enfants sont importants pour la santé de mon enfant; (2) Les vaccins pour enfants sont efficaces; (3) Faire vacciner mon enfant est important pour la santé des autres au sein de ma communauté; (4) Tous les vaccins pour enfants offerts par le programme du gouvernement dans ma communauté sont bénéfiques; (5) Les nouveaux vaccins sont plus porteurs de risques que les anciens; (6) Les renseignements que je reçois concernant les vaccins de la part du programme de vaccination sont fiables et digne de confiance; (7) Faire vacciner mon est un bon moyen de le protéger contre les maladies; (8) Généralement, je fais ce que mon médecin ou professionnel de la santé recommande concernant la vaccination de mon enfant; (9) Je suis concerné par les effets indésirables graves des vaccins |

* Items 1, 2, 3, 4, 6, 7, and 8 were reverse-coded

**Results**

*Descriptive Results*

Table 3 presents the descriptive statistics for the primary model. The mean score in each column was derived from 30 iterations (different sessions) of the same condition. We calculated the mean of each question and the mean across all nine questions. In addition, the VHS scale was divided into two factors including 'lack of confidence' (Conf) and

'risk' (Risk), as described previously and we report the mean scores for them too. The bottom row of Table 3 refers to the mean value of each column. We notice that 'risk' was a more prominent factor for vaccine hesitancy (overall mean = 2.49) than 'lack of confidence' (overall mean = 1.35).

Table 3. Means (and standard deviations) across iterations for Vaccine Hesitancy Scale questions using ChatGPT

|  | Q1 | Q2 | Q3 | Q4 | Q5 | Q6 | Q7 | Q8 | Q9 | VHS mean | Conf mean | Risk mean |
|---|---|---|---|---|---|---|---|---|---|---|---|---|
| **English** | 1.43 (0.50) | 1.63 (0.49) | 1.67 (0.48) | 1.93 (0.25) | 2.10 (0.55) | 1.97 (0.18) | 1.50 (0.51) | 1.77 (0.43) | 3.07 (0.69) | **1.90 (0.23)** | **1.70 (0.27)** | **2.58 (0.51)** |
| **Spanish** | 1 (0) | 1 (0) | 1 (0) | 1.47 (0.51) | 2.20 (0.55) | 1.53 (0.57) | 1 (0) | 1.07 (0.25) | 2.37 (0.61) | **1.40 (0.13)** | **1.15 (0.13)** | **2.28 (0.41)** |
| **French** | 1.03 (0.18) | 1 (0) | 1.03 (0.18) | 1.67 (0.71) | 2.20 (0.41) | 1.43 (0.50) | 1 (0) | 1.13 (0.35) | 3.03 (0.81) | **1.50 (0.17)** | **1.19 (0.17)** | **2.62 (0.47)** |
|  | **1.16 (0.36)** | **1.21 (0.41)** | **1.23 (0.43)** | **1.69 (0.55)** | **2.17 (0.50)** | **1.64 (0.50)** | **1.17 (0.37)** | **1.32 (0.47)** | **2.82 (0.77)** | **1.60 (0.28)** | **1.35 (0.32)** | **2.49 (0.48)** |

*Results for RQ1: Comparing Human and ChatGPT Responses*

In Table 4, we summarize a comparison of means by language between datasets from a prior study (Shapiro et al.[1]) and those observed for ChatGPT. Overall, the results show that the mean values of all questions regarding the VHS scale were lower in ChatGPT responses than in human responses, regardless of language condition (EN or FR). As seen in Table 4, the averages of ChatGPT responses in English ($M = 1.90$) and French ($M = 1.50$) were smaller than those of human responses in English ($M = 2.21$) and French ($M = 2.27$).

Table 4. Vaccine hesitancy averages (and std. dev.) reported in Shapiro et al. and those obtained from ChatGPT

|  | Human-Generated Results | | | ChatGPT-Generated Results | | | |
|---|---|---|---|---|---|---|---|
|  | English | French | Both Languages | English | Spanish | French | All Languages |
| **All Questions** | 2.21* | 2.27* | 2.23^ | 1.90 (0.23) | 1.40 (0.13) | 1.50 (0.17) | 1.60 (0.28) |
| **Lack of Confidence** | 1.97 (0.72) | 2.03 (0.72) | 1.98 (0.72) | 1.70 (0.27) | 1.15 (0.13) | 1.19 (0.17) | 1.35 (0.32) |
| **Risk** | 3.06 (0.96) | 3.12 (0.89) | 3.07 (0.95) | 2.58 (0.51) | 2.28 (0.41) | 2.62 (0.47) | 2.49 (0.48) |

* Approximated from Shapiro et al.[1] (Table 6). Not directly reported in Shapiro et al.[1]; this value is calculated for each language as a weighted average of 'Risk' and 'Lack of Confidence' factors.
^ Approximated from Shapiro et al.[1] (Supplementary document, Table A3, items 1 through 9 only). The value is calculated as the mean of all nine-item means.

Furthermore, this trend was persistent across the VHS factors ('lack of confidence' and 'risk'). With regard to each factor of the VHS, ChatGPT responses were less hesitant in both languages, compared to human responses. In specific, for the factor 'lack of confidence', there were lower mean values of ChatGPT responses in English ($M = 1.70$, $SD = 0.27$) and French ($M = 1.19$, $SD = 0.17$), compared to the means of human responses in English ($M = 1.97$, $SD = 0.72$) and French ($M = 2.03$, $SD = 0.72$). In the context of the 'risk' factor, the averages of ChatGPT responses in English ($M = 2.58$, $SD = 0.51$) and French ($M = 2.62$, $SD = 0.47$) were also smaller than the averages of human responses in

English ($M = 3.06$, $SD = 0.96$) and French ($M = 3.12$, $SD = 0.89$). Taken together, these results indicate that in both languages (EN and FR) and across both the VHS factors, the average hesitancy values of ChatGPT were lower than those of human participants in Shapiro et al.[1].

*Results for RQ2: Comparing Responses in English, Spanish, and French*

We compare the average hesitancy scores in responses coming from ChatGPT in different languages (EN, ES, and FR) (See first row in Table 4). The average hesitancy levels indicate most hesitancy in English (mean=1.90) followed by French (mean= 1.50) and then Spanish (mean= 1.40). A one-way ANOVA test over these values indicated that these differences were significant across language groups [$F(2, 87) = 62.24$, $p < .001$]. Next, we zoom into the factors of the VHS scale ('lack of confidence' and 'risk'; shown in rows 2 and 3 of Table 4). Based on a one-way ANOVA test based on languages, we find the differences between languages for the 'lack of confidence' factor to be statistically significant [$F(2, 87) = 72.23$, $p < .001$]. A similar statistically significant difference was observed for the VHS 'risk' factor [$F(2, 87) = 4.698$, $p = .0115$].

Overall, these results show significant differences in ChatGPT's VHS responses across languages, with English responses, on average, showing the most hesitancy and Spanish the least.

*Results for RQ3: Impact of Variations in the Model*

Table 5 summarizes the results of the primary model and three variations, each differing by a single aspect: temperature, model type, or number of questions per session. One-way ANOVA analysis indicates that language plays a significant role in predicting the average hesitancy levels in each of the four models (primary + three variations). Similarly, a two-way ANOVA reveals significant differences across factors and languages in each of the four configurations. This corroborates RQ2 results about significant differences in hesitancy levels across languages.

Next, we conduct post-hoc pairwise comparison tests with Bonferroni correction (alpha=0.167) for significance in the observed differences between specific pairs of languages (EN vs. ES, EN vs. FR, FR vs. ES). Comparisons on the mean values of all questions show that English responses had higher vaccine hesitancy than Spanish (respectively French) across all four configurations. The differences between French and Spanish were less pronounced. While Spanish responses consistently indicated the least hesitation, the differences with French were significant in only two of the four configurations. Similarly, pairwise comparison tests (with Bonferroni correction) were conducted between languages separately for the two factors ('lack of confidence' and 'risk'). For 'lack of confidence', vaccine hesitancy was consistently higher in English than in Spanish (respectively French) in all configurations. However, the differences between Spanish and French were not significant for all four configurations. Regarding the 'risk' factor, the highest vaccine hesitancy varied between English and French, but hesitancy for Spanish was consistently the lowest. The differences between Spanish and English (respectively, French) were significant in three of the four configurations but the differences between English and French were consistently not significant.

Overall, the results indicate a consistent trend of differences between the three languages in terms of average hesitancy. However, the results varied when zooming in on individual factors.

**Discussion**

In this study, we analyzed the quality and equity of ChatGPT-generated responses to vaccine hesitancy by using an existing vaccine hesitancy scale. First, we compared ChatGPT responses with human responses reported by Shapiro et al.[1]. Our comparative analysis shows that ChatGPT responses were on average less hesitant than the levels reported for human respondents in the past (RQ1). This trend was consistent across both comparable languages (EN vs. FR) and the two factors of the VHS ('lack of confidence' vs. 'risk'). We also compared ChatGPT responses in three different languages (EN, ES, and FR) with respect to vaccine hesitancy. Our inter-language analysis revealed that there are significant differences in ChatGPT responses across languages (RQ2). On average, vaccine hesitancy was highest in English responses and lowest in Spanish responses. Finally, we identified differences by the VHS factors and various model variations (e.g., temperature, model type, and session window). Our results showed that average differences in ChatGPT responses across languages were stable regardless of variations in the technical parameters (RQ3). However, these differences did not remain as consistent when we analyzed them at a factor level. For instance, while English responses consistently indicated more hesitancy than French in terms of 'lack of confidence', the differences between English and French were not significant in terms of 'risk'.

**Table 5.** Summary of results obtained for the primary model and model variations in VHS responses

|  | Primary Model GPT-3.5-Turbo | Model Variations | | |
|---|---|---|---|---|
|  |  | Change of Temperature GPT-3.5-Turbo | Change of Model Type GPT-4 | Change of #Q in Session GPT-3.5-Turbo |
| **Temperature** | default | 0.5 | default | default |
| **# Question in session** | 9 per session | 9 per session | 9 per session | one-by-one |
| **Two-way: Factor x Lang** | sig | sig | sig | sig |
| **One-way: Lang** | sig | sig | sig | sig |
| **Language means** | | | | |
| pairwise: English-Spanish | sig | sig | sig | sig |
| pairwise: Spanish-French | sig | NOT sig | NOT sig | sig |
| pairwise: English-French | sig | sig | sig | sig |
| mean English | 1.896 | 1.985 | 1.915 | 2.152 |
| mean Spanish | 1.404 | 1.407 | 1.441 | 1.674 |
| mean French | 1.504 | 1.441 | 1.489 | 1.756 |
|  | en >> fr >> es | en >> fr ~ es | en >> fr ~ es | en >> fr >> es |
| **One-way: LackConf** | sig | sig | sig | sig |
| **One-way: Risk** | sig | sig | sig | sig |
| **Factor: Lack of Confidence** | | | | |
| pairwise: English-Spanish | sig | sig | sig | sig |
| pairwise: Spanish-French | NOT sig | NOT sig | NOT sig | NOT sig |
| pairwise: English-French | sig | sig | sig | sig |
| mean English | 1.700 | 1.762 | 1.700 | 1.914 |
| mean Spanish | 1.152 | 1.186 | 1.195 | 1.343 |
| mean French | 1.186 | 1.110 | 1.186 | 1.376 |
|  | en >> fr ~ es | en >> es ~ fr | en >> es ~ fr | en >> fr ~ es |
| **Factor: Risk** | | | | |
| pairwise: English-Spanish | sig | sig | sig | NOT sig |
| pairwise: Spanish-French | sig | sig | NOT sig | sig |
| pairwise: English-French | NOT sig | NOT sig | NOT sig | NOT sig |
| mean English | 2.583 | 2.767 | 2.667 | 2.983 |
| mean Spanish | 2.283 | 2.183 | 2.300 | 2.833 |
| mean French | 2.617 | 2.600 | 2.550 | 3.083 |
|  | fr ~ en >> es | en ~ fr >> es | en ~ fr ~es | fr ~ en ~es |

These findings have implications for health informatics research and practice. First, our results suggest that ChatGPT is a useful novel source of information on vaccination, as it adopts a scientifically aligned, vaccination-affirming tone that may increase vaccine acceptance and confidence by the public. This is consistent with previous studies that found that ChatGPT can generate accurate and relevant answers to health-related questions in some contexts[31,32]. This is also important since ChatGPT may have a persuasive effect on users given that it uses natural language and conversational strategies to engage them and address their concerns. For example, recent work exploring ChatGPT's responses to 13 cancer myths and misconceptions found that 96.9% were accurate and did not provide misinformation or harmful information; responses for all questions were consistent in five (5) runs of the questions[32]. ChatGPT, therefore, could be a valuable tool for health communication and education, especially in contexts where there is limited access to health professionals or credible information sources.

Second, our results highlight the importance of considering language diversity and health equity when evaluating and designing web information systems for health-related topics. Although generally less hesitant than responses reported by Shapiro et al.[1]., we found that ChatGPT responses varied significantly across languages, with more hesitancy in English responses and less hesitancy in Spanish responses. This may reflect the differences in the training data and the linguistic features of each language, as well as the differences in the socio-cultural contexts and attitudes towards vaccination in different regions from which the training data may have been sourced. For instance, Coleman[33] reports that Spain had a higher level of vaccination than France and both were higher than the rate in the USA. Similarly, a recent systematic review on COVID-19 vaccine acceptance in Latin America notes that vaccine intentions were relatively higher than or similar to studies conducted in European and Asian countries[34]. If web sources from these countries impact the training data (in Spanish, French, and English respectively), they could influence the relative levels of hesitation demonstrated in ChatGPT responses. For example, English responses may be more hesitant because they are influenced by anti-vaccine movements and misinformation that are prevalent in some English-speaking countries[35]. Spanish responses may be less hesitant because they are influenced by the high vaccination coverage and trust in health authorities that are common in some Spanish-speaking countries[36,37,38]. These differences may have implications for the effectiveness and equity of ChatGPT as a health information source, as it may not equally address the needs and preferences of users from different linguistic and cultural backgrounds.

Our findings also have public health implications. They provide public health researchers and healthcare providers with a new methodological toolkit to understand potential pitfalls in how ChatGPT discusses important health topics in different languages. The toolkit could be used by other researchers to explore similar questions with ChatGPT. The insights obtained can also help healthcare providers anticipate vaccine hesitancy-related questions (e.g., those showing most variations in answers across languages or configurations) that the laypersons may bring to patient-provider encounters. From a policy perspective, this work contributes to a growing body of literature attempting to understand the implications new LLM and conversational AI technologies can have on the type and quality of information laypersons receive. A lack of transparency in these processes may put certain populations at higher susceptibility to health misinformation and this work marks an early attempt at making them less opaque.

The findings can also have implications for the designers of conversational AI algorithms. The current results indicate that despite a uniform inter-language interface, the training and result-generation processes in ChatGPT likely work in silos for different languages. Newer approaches that combine the data available in different languages to provide holistic and equitable responses for public health-related queries could be a major step toward equitable and reliable use of such technology. Some of the potential ways to support this include ensuring that culturally and linguistically diverse data are part of the training process, creating self-assessment/ "regularization" mechanisms that ensure that responses to the same query in different languages are similar, and building uniform "wrapper" widgets that combine responses generated in different languages into a common holistic response.

*Limitations*

This study has some limitations that should be acknowledged. First, we used a single scale (VHS) to measure vaccine hesitancy, which may not capture all dimensions or nuances of this complex phenomenon. Future studies could use other scales or methods to assess vaccine hesitancy from different perspectives. Second, we compared the human responses as reported for a Canadian population in Shapiro et al.[1] in 2016 with the ChatGPT-generated responses in the US in 2023. Severe public health emergencies between the two periods, such as the COVID-19 pandemic, and the different levels of French language predominance between the two regions may affect the differences between the two datasets. We acknowledge that the change in context does not allow for the numbers to be directly comparable but rather as a way to interpret the ChatGPT scores. We acknowledge that the translation of the VHS scale to Spanish, validated by a bilingual, bicultural team member who is an expert in public health communication, has not been validated externally. We used a limited number of iterations (n=30) to obtain the average results for each language, model, and parameter combination, which may not be sufficient to detect subtle differences or trends. Future studies could use larger or more diverse samples to increase the generalizability and statistical power of the results. We hope that our publicly released code and dataset can help in that process. Lastly, we do not tease apart the differences between language and culture in this work. We maintain focus on comparison across the languages of interaction with ChatGPT. Future work could combine survey scales with more detailed conversation analysis to add more nuance to the comparative analysis.

**Conclusion**

In conclusion, this study provides a novel and comprehensive evaluation of ChatGPT responses regarding vaccine hesitancy by using a validated scale. We found that ChatGPT responses indicated less hesitancy than human respondents in North America in the past and that there were significant differences across languages (EN, ES, and FR), with the most hesitant responses being observed in English and the least hesitant responses in Spanish. We also found that ChatGPT responses were largely consistent across different models and parameters but showed some variations in the factors of vaccine competency and risk. These findings have implications for health informatics researchers interested in evaluating and improving the quality and equity of web information sources for health. However, we also acknowledge the limitations of the study and the challenges of using conversational AI technologies for health information seeking. Future research should explore new approaches that combine the data available in different languages to provide holistic and equitable responses to public health-related queries. This could be a major step toward widespread and reliable use of such technology for supporting public health.